\documentstyle[aps,epsf,preprint,epsfig]{revtex}
\def\beq{\begin{equation}}
\def\eeq{\end{equation}}
\def\beqa{\begin{eqnarray}}
\def\eeqa{\end{eqnarray}}
\def\l{\left}
\def\r{\right}
\def\bdi{\begin{displaymath}}
\def\edi{\end{displaymath}}

\begin{document}

\title{Force-induced unfolding of a homopolymer on fractal lattice:
exact results vs. mean field predictions}

\author{ 
Davide
Marenduzzo$^1$, Amos Maritan$^1$ $^2$  and 
 Flavio Seno$^3$}
\vskip 0.3cm

\address{$1$ International School for Advanced Studies,
 INFM, Via Beirut 2, 34014 Trieste Italy,\\
$2$ Abdus Salam International
Center for Theoretical Physics, Via Beirut 2, 34014 Trieste Italy,\\
$3$ INFM - Dipartimento di Fisica ``G. Galilei'' , Via
Marzolo 8, 35100 Padova, Italy}


\maketitle

\begin{abstract}
We study the force-induced unfolding of a homopolymer 
on the three dimensional Sierpinski gasket. The polymer is subject to a
contact energy between nearest neighbour sites not consecutive along the chain
and to a stretching force.
The hierarchical nature of the lattice we consider allows for an exact 
treatment which yields the phase diagram and
the critical behaviour. We show that for this model mean field predictions are
not correct, in particular in the exact phase diagram 
there is {\em not} a low temperature reentrance and we find 
that the force induced unfolding transition below the theta temperature
is second order.
\end{abstract}

\noindent{64.60.Cn, 87.15-v, 87.15.He, 05.10.Cc}
\newpage

The recent  development of single molecule techniques have given
experimentalists the opportunity to grab with suitable handles
and mechanically manipulate, by means of optical tweezers\cite{optical}
or cantilevers such as atomic force microscopes\cite{AFM}, 
proteins, molecular motors or DNA molecules. 
In this way it has become possible to measure or exert on these molecules
forces in the piconewton range. In particular the 
effect of a stress on the giant
molecule titin has been studied in\cite{titin1,titin2,titin3} 
where it was found that
a force induced unfolding transition takes place between a compact native
like state and an extended state. The presence of strong hysteresis 
together with the rather abrupt jumps observed in the force vs. elongation 
curves suggest that the unfolding is a first
order phase  transition. 
On the theoretical side, many simple models have been proposed
to make contact with
these experiments (see e.g. \cite{shak1,shak2,grassberger,hoang,thir,zhou} 
for the stretching
of proteins and homopolymers, \cite{somen,nelson,rientranza,jpu} 
for DNA unzipping and \cite{andreamontanari,gerland,cocco} for RNA unzipping). 
In particular,  in Refs.\cite{shak1,shak2} the authors
study the force-induced unfolding of a homopolymer and 
a heteropolymer in the mean field
approximation and find that in both cases the critical line separating the
globule from the coil is re-entrant at low temperatures. This is analogous
to what had been found  for the phase boundary in the phase diagram 
valid for the DNA unzipping, where it has been proved
exactly that in the presence of a pulling force mean field is
correct\cite{rientranza}. 

When self-avoidance is incorporated in the models, exact results
are rare and mean field treatments become popular.
In this work, we study exactly the force-induced unfolding of a self-avoiding
walk (SAW) on a fractal lattice, the three-dimensional Sierpinski Gasket (SG)
\cite{dhar1,dhar2}. This gasket has a fractal dimension $D_f=2$, and this 
renders the
system interesting from the theoretical point of view because we are
below the upper critical dimension for theta collapse\cite{degennes}. 
Mean field
theory would predict a reentrance here because the hamiltonian compact walks
in this lattice have nonzero entropy as compared to the
zero entropy of the completely stretched coil (same argument as in
\cite{rientranza}).
However, the exact critical line that we find in the SG
shows no reentrance. Similarly,
by comparing the free energies 
of the globule and that of the coil one would naively
expect a first order transition, whereas the exact calculation yields
a second order transition.
Though we cannot conclude from this calculation that also on 
Euclidean lattices the situation will be analogous, 
we feel that this calculation
should give a warning that the mean field prediction need not be correct.
The presence or not of reentrance on Euclidean lattices 
together with the nature of transition is thus an intriguing question 
which deserves further work.

We consider a SAW on the three-dimensional SG (3DSG)\cite{dhar1,dhar2}, 
a hierarchical 
lattice with ramification number $4$ and $D_f=2$ (Fig. 1). 
We study the combined effect of a force $\vec f$
which stretches the polymer along one edge of the SG 
(Fig. 1) together with a compacting self-attractive term, obtaining by assigning
a weight $\exp(\beta\epsilon)$ ($\epsilon>0$) 
every time two non consecutive sites of the 
SAW are nearest neighbour in the 3DSG. We call 
$\beta\equiv T^{-1}$ the inverse temperature.
To describe the effect of the stretching force,  
we give each step of the walk an orientation and a weight 
$\exp(\beta f\Delta a)$, where $f\equiv |\vec f|$ is the modulus of $\vec f$
and  $\Delta a$ is the projection of the 
oriented step along $\vec f$. 

The calculation of the partition function and other thermodynamic
quantities involves the evaluation of twenty-five generating functions
(Fig. 2). Twelve generating functions involve contributions
of a SAW which enters in the $n$-th order SG at one vertex and
goes out at one other vertex. Other twelve generating functions
arise when an (oriented) SAW starts from a vertex, goes out from 
another one and then re-enters the $n$-th order SG at a later 
stage. The last generating function is the void.
The one-leg generating functions are labelled $A_i$, $i=
1,\ldots,12$, and the  two-leg contributions 
$B_i$, $i=1,\ldots,12$. One can write recursion
relations for the generating functions by exactly enumerating all walks 
in the SG on the computer: one has to sum up the contributions
of SAWs at the $n-$th order SG in order to generate SAWs in the
$n+1$-th order SG. 

The initial conditions for the generating functions are:
\beqa\label{generating_function}
A_{1} & = & zy^{-1}\ \ \ \ \ \ \ \ \
A_{2,3,4,8} =  zy^{-1/2} \\ \nonumber
A_{5,9,10,11} &  = &  zy^{1/2} \ \ \ \ \ \ \ \ 
A_{6,7}  =  z \ \ \ \ \ \ \ \ \ \ \ \ \ 
A_{12}  =  zy\\ \nonumber
B_{1,2,3,6} & = & z^2 w^4 y^{-1}\ \ \ \
B_{4,5,8,9}  =  z^2 w^4\ \ \ \
B_{7,10,11,12} =   z^2 w^4 y 
\eeqa
where $z$ is the SAW step fugacity, $y\equiv\exp{\l(\beta f\r)}$, and
$w\equiv\exp{\l(\beta\epsilon\r)}$, is the weight 
responsible for the theta collapse in absence of force.

Note that, as is apparent from the above equations, we have adopted
the convention that interactions are restricted to
sites within the first order SG and moreover that after one 
SAW has touched one vertex in the SG it is obliged to exit from
the SG at that order\cite{dhar2,dhar3,knezevic}.
 It has been proved that this approach is equivalent to the more 
general approach as regards universality of the phase transition
\cite{knezevic}.

When there is no pulling force the recursion relations simplify
because:
\beqa
A_i  \equiv  A \qquad\qquad \ \ \ \ \ 
B_i  \equiv  B \qquad\qquad \forall i=1,\ldots,12 ,
\eeqa
so one gets\cite{dhar2}: 
\beqa\label{rg_zero_force}
A'& = & A^2 + 2 A^3 + 2 A^4 + 4 A^3B + 6 A^2B^2,\\
B'& = & A^4 + 4 A^3B +22 B^4\nonumber.
\eeqa

At non-zero $\vec f$, the equations retain the same structure but 
every $B$ and every $A$ has to be labelled by the appropriate 
number as found in the exact enumeration.
To solve the model, we proceed as follows. First,
we must find the fixed points of the recursion relations: in general, 
for every phase or critical point (line) there is a fixed point. 
Then, for every $f$ and $T$ fixed, we have to find the critical step fugacity, 
$z_c(\beta\epsilon,\beta f)$ such that 
the flux defined by the twenty-five recursion 
relations and initial conditions will bring the system in the fixed point 
corresponding to those values of $f$ and $T$. This critical step fugacity 
$z_c(\beta\epsilon,\beta f)$ allows us to have a numerical expression for all 
the quantities we are interested in, namely, the free energy $F(\beta\epsilon,
\beta f)\equiv\frac{1}{\beta}\log{(z_c(\beta\epsilon,\beta f))}$,
 the average elongation along the direction of the pulling force, 
$\langle x\rangle(\beta\epsilon,\beta f)$ and the average number of contacts, 
$\langle n\rangle(\beta\epsilon,\beta f)$, found by calculating the 
appropriate derivative of $F(\beta\epsilon,\beta f)$.

From the linearized recursion equations , 
it is possible to get the critical exponents of the
phase transition. Every eigenvalue $\lambda$ (there are at most as many 
such eigenvalues as there are recursion relations) , of the linearized flux 
at given $f$ and $T$ defines two critical exponents, $Y$ and $\nu$, through:
\beq
\lambda\equiv 2^{Y}\equiv 2^{\frac{1}{\nu}}
\eeq
where $\nu\equiv\frac{1}{Y}$ is defined in terms of the average squared
elongation $\langle x^2\rangle$ as $\langle x^2\rangle
\sim N^{2\nu}$\cite{degennes} 
for large values of the number of steps in the SAW, $N$.

Performing this analysis for our model, we find six different fixed points.
At zero force, we re-obtain the fixed points given in \cite{dhar2}.
In particular, for $T<T_{\theta}\equiv \frac{2}{\log(3)}$,
the theta collapse temperature in the absence of force,  there is
a zero force compact fixed point for the recursion relations:
\beqa\label{compact_zero_force}
A_i  =  0 \qquad\qquad \ \ B_i  =  22^{-1/3}\equiv B^* \qquad\qquad 
\ \ \forall i=1,\ldots,12.
\eeqa
For $T>T_{\theta}$, i.e. in the swollen phase,
 the fixed point that is approached is:
\beqa\label{swollen}
A_i  \simeq 0.4294\ldots  \ \ \ \ \
B_i  \simeq 0.04998\ldots \qquad \qquad \ \ \forall i=1,\ldots,12.
\eeqa
Just at criticality at the theta temperature, $T=T_{\theta}$, 
the fixed point is:
\beqa
A_i =  1/3 \qquad\qquad \ \ \
B_i =  1/3 \qquad\qquad \ \ \ \qquad \forall i=1,\ldots,12.
\eeqa
The corresponding critical indeces $\nu$ (corresponding to the
largest eigenvelues, $\lambda_1$, in every one of the three regimes) 
are $1/2$, $0.5294\ldots$ and
$0.7294\ldots$ respectively in the collapsed phase, at the theta
point and in the swollen phase.

At $T<T_{\theta}$, there exists a critical line $f=f_c(T)$
separating a compact from an open phase. When $f<f_c(T)$ the
recursion relations display another compact fixed point:
\beqa\label{compact_nonzero_force}
A_{i} & = & 0 \qquad\qquad \forall i=1,\ldots,12 \\
B_{4,5,8,9} & = & 22^{-1/3} \ \ \
B_{1,2,3,6}  =  0 \ \ \
B_{7,10,11,12} = +\infty \nonumber 
\eeqa
The divergence arises from the pulling force and one can convince
that the two leg generating functions are diverging (vanishing) for
$L\to\infty$ ($L$ is the system size, and one has $L=2^{n-1}$ at the
$n$-th level of iteration in the construction of the 3DSG) 
as $y^L$ ($y^{-L}$). Consequently, we can
eliminate the divergences if e.g. we multiply the diverging functions
for $B_{1}$, and the vanishing ones for $B_7$, and then take the square root
of the result: if we do this all the
two leg generating functions converge to the fixed point $22^{-1/3}$ as in
the zero force compact phase. 
When the SAW is in the open phase ($f>f_c(T)$), the fixed point is:
\beqa\label{open}
A_{i} & = & 0 \qquad\qquad \forall i=1,\ldots,11; \ \ \
A_{12} = 1\\
B_{i} & = & 0 \qquad\qquad \forall i=1,\ldots,12  \nonumber
\eeqa
Finally, when the force is exacly tuned at the critical line, the
fixed point which is approached is:
\beqa\label{critical_force}
A_{i} & = &  0 \  \  \ \forall i=1,\ldots,11; \ \
A_{12} = \frac{1}{1+6 * 22^{-2/3}}\sim 0.5668\ldots\equiv A^*\\
B_{4,5,8,9} & = & 22^{-1/3} \ \ \
B_{1,2,3,6}  =  0 \ \ \
B_{7,10,11,12}  =  +\infty \nonumber
\eeqa
again the square root of the 
products of one diverging two leg function times a vanishing two
leg function is $22^{-1/3}$. We will see later that it is possible to
justify the value found numerically also for the nonzero force system. 
The exponent $\nu$ is equal to $1/2$ for $f\le f_c(T)$ and is $1$ in the
open phase.

In practice, by taking advantage of the knowledge of the fixed points,
one can devise a convenient way to calculate numerically the critical
force as a function of temperature.  Once the critical step fugacity
$z_c(\beta\epsilon,\beta f=0) $, of the compact phase at a fixed
temperature $T$, is known one can further fix $z$ to this value, and
then tune $f$ in order to pass from the compact fixed point to the
critical line fixed point (within the precision allowed by the
computer).  In order to have a sufficient accuracy in these
calculations, it proved necessary to use a computer with quadruple
precision.

The phase diagram obtained numerically is shown in Fig. 3. Two remarks are in
order. First, we notice that there is no re-entrance in the critical line, the
slope at $T=0$ being approximately $-0.05$. This is
at variance with the prediction doable on the basis of 
mean field-like treatments (as those done e.g. in  Refs.\cite{shak1,shak2}, 
see also below). Second, the behaviour of the critical line near 
$T_{\theta}$ is:
\beq\label{theta_behaviour}
f_c(T)\sim (T_{\theta}-T)^{a=0.87\pm 0.01},
\eeq
in agreement with the prediction $a=\frac{\nu_{\theta}}{\phi}=0.868$, where
$\nu_{\theta}$ is the critical exponent of the end-to-end distance at the
theta temperature and $\phi$ is the theta transition crossover
exponent\cite{degennes}
(see below for an argument, leading to the behaviour in 
Eq. \ref{theta_behaviour}).

We report in Fig. 4 the plot of $\langle x\rangle$ vs. $f$ at
$T=0.35$. This figure supports the hypotesis 
that the unfolding transition for the homopolymeric SAW on the 3DSG is second
order. This is in agreement with the argument based on the recursion
relations given below.
Our belief supported by the exact numerics is that the transition is second 
order for any nonzero $T$ and is first order only at $T=0$.

We now present some arguments to interpret our results (based on the
renormalization group (RG) flux).
Even though to write explicitly the equations
and to cope numerically with them it was 
necessary to put the force in step by step, it is equivalent to 
evolve the $f=0$ generating functions (Eq. \ref{rg_zero_force}) and then
put in the dependence on $\vec f$ at every iteration
by multiplying the generating functions times suitable powers of $y$.
If we do this, we can exploit the symmetries of the problem which make so that
before multiplying the generating functions times $y$ dependent term, 
all one leg and two leg diagrams are separately equivalent. By
noting the structure of the nonzero force fixed points, one can argue that
the one leg generating functions must vanish in the
compact phase like $A\sim C y_c^{-L}$, 
where $y_c\equiv\exp{\l(\beta f_c(T)\r)}$
and $C$ is a constant, for now undetermined. 
By matching the exponentials in Eq.\ref{rg_zero_force}, we obtain that the two
equations, when $T\ll T_{\theta}$ ($T\to 0$), can be approximated by:
\beqa\label{rg_approx}
A' =  A^2 + 6A^2B^2,\qquad \qquad
B' =  22 B^4.
\eeqa
The flux corresponding to these RG equations is shown in Fig. 5.
The non-trivial fixed point, $(A^*,B^*)$  with both $A^*$ and $B^*$ non-zero,
is obtained with $B^*=22^{-1/3}$ and $A^*=\frac{1}{1+6\l(B^*\r)^2}$.
Thus $A\sim C y_c^{-L}$ as $L\to\infty$, with $C=A^*$. 
When $T\to 0$, we notice that the fixed points 
can be approached at once, i.e. since 
the initial condition, also in the original (not simplified) recursion 
relations, because the terms that should evolve to $0$ 
have an initial value which {\em is} zero for $T\to 0$. 
As a consequence, as $T\to 0$, the phase boundary of Fig. 3 is found by 
matching the initial conditions with
the fixed point values for $A_{12}$ and any $B$.
Thus the critical line is found by solving this system:
\beqa
z_c & = & \exp{\l(-2\beta\epsilon\r)} \l(B^*\r)^{1/2},\\
z_c & = &  \exp{\l(-\beta f\r)} A^* \nonumber,
\eeqa
where $B^*$ and $A^*$  are the nonzero 
fixed points of Eq. \ref{rg_approx},
corresponding to the critical
force (such that $y^L A\sim 1$ for large $L$, 
see Eq.\ref{critical_force}). 
The critical line for $T\to 0$ is:
\beq\label{limit_Tto0}
f_c(T) \sim 2\epsilon + T \log{\l(\frac{A^*}{(B^*)^{1/2}}\r)} \sim 2\epsilon -
0.0525\ldots T,
\eeq
so that $f_c(T)$ starts with negative slope as found numerically.
For $f>f_c(T)$, the fixed point to be reached in the $(A,B)$ plane
of Fig. 5 is $(1,0)$. We observe that for a small deviation from
$A^*$ the RG flux takes the SAW to the open fixed point by moving along
 the line $B(A)$, which we can find in the neighbourhood of $(A^*,B^*)$ by
requiring that it be a fixed line under the flux defined by the 
recursion equations.
We require that $B(A)\sim B^*+c(\delta A)^{\alpha}$ in the neighbourhood
of $A^*$. We thus need to solve the system:
\beqa
A' & = & \left[A^*+(\delta A)\right]^2 
\left[1+6\left[B^*+c(\delta A)^{\alpha}\right]^2\right]\\
B(A') & = & 22 \left[B^*+c(\delta A)^{\alpha}\right]^4\nonumber
\eeqa
in the unknown quantities $c$ and $\alpha$.
We obtain $c = -1/{12{A^*}^3B^{*}}\sim -1.2826\ldots$ and 
$\alpha=2$ so that the fixed line $B(A)$ approaches smoothly the fixed point
at critical force.
By inserting these values in the expression $B(A)\sim B^*+c(\delta A)^{\alpha}$
 together with the ansatz: 
\beq
z_c(\beta f_c,\beta\epsilon)-z_c(\beta f,\beta\epsilon)\sim (f-f_c)^{\gamma},
\eeq
one gets $\gamma=2$ which  implies $\langle x\rangle\sim (f-f_c)$ 
for $f\stackrel{>}{\sim} f_c$ and the transition is second order. 
This argument is strictly valid for low $T$. However it is unlikely 
that the order of the transition could change along the phase boundary
and this is also confirmed by our numerics. The situation at the point $T=0$ 
is somewhat  special: the entropy
vanishes and balancing the energetic terms gives a first order transition.

One should notice the importance
of the term $6 A^2 B^2$. Let us
think of a mathematical simplified model in which the relevant
equations are:
\beq\label{meanfield_recursion}
A'  =  A^2,\qquad \qquad 
B'  =  22 B^4 ,
\eeq
in which the mixed term (physically due to stretched walks that still 
make an extensive number of contacts) is suppressed. This corresponds 
to balancing the free energy of a stretched coil with that of a 
compact globule. In this simplified treatment
the re-entrance is present, as can be expected from
naife estimates of the ground state entropies of the stretched 
and compact state, and the transition is first order.

Finally, we argue that the exponent $a$ in Eq.\ref{theta_behaviour} is given 
by $a=\frac{\nu_{\theta}}{\phi}$ as anticipated above. Near $T_{\theta}$  at
zero force the free energy behaves as $F(T)\sim (T_{\theta}-T)^{2-\alpha=
\frac{1}{\phi}}$. For $f\stackrel{>}{\sim} 0$, on the other hand, 
at $T=T_{\theta}$,
the free energy behaves as $F(f)\sim f^{\frac{1}{\nu_{\theta}}}$. 
Consequently, we get $a=\frac{\nu_{\theta}}{\phi}$ in Eq. 
\ref{theta_behaviour}.
This is in agreement with the result found in $d=3$ in Ref.\cite{grassberger}
where the exponents both take their mean field values ($1/2$), and also with
the mean field analysis in Ref.\cite{shak1,shak2} in which $\phi=1$ and
consequently $a=\nu$ ($1/2$ in the ideal case treated in \cite{shak1,shak2}).

In conclusion, we presented an exact calculation of the phase 
diagram of a SAW in presence of a compacting contact
energy and a stretching force. We deem it is interesting because it can be
analyzed exactly. A mean field like treatment 
 gives a reentrant boundary and a first
order transition. Both these predictions are not confirmed by the 
exact treatment, which gives a critical line with a zero temperature
negative slope and a second order transition. This has been explained in a 
simple way by analyzing a simplified version of the recursion relations,
analogous to the real space RG equations.
Whether or not the critical line in the hypercubic lattice shows reentrance  
is therefore not yet clear and appears to be an intriguing question. 
The 3DSG has $D_f=2$ and so the most natural comparison is with the 
Monte-Carlo simulations
performed in Ref.\cite{grassberger}, which indeed gives a second order
transition, even though we cannot be sure that the order of the transition
is the same in the two-dimensional real and fractal lattices. The behaviour of
the critical force near the theta point has been found and a general 
argument, valid also for
hypercubic lattices, has been given in agreement with the result found here
and also with the result found in $d=3$ in\cite{grassberger}. 

This work was supported by cofin2001.


\begin{figure}
\begin{center}
\epsfxsize=3.2in
\centerline{\epsfbox{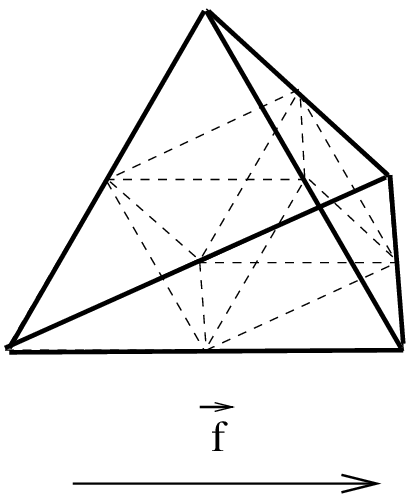}}
\end{center}
\caption{In this figure we show the 3DSG at the second stage of iteration 
construction. Four tetrahedra (dashed lines) at the first order of 
iteration are put together in order to form the 
tetrahedron which constitutes the 3DSG at the second iteration 
(bold lines). The stretching force $\vec f$ acts along one of the edges 
of the 3DSG and is also shown in the figure.}
\end{figure}

\begin{figure}
\begin{center}
\epsfxsize=6.in
\epsfbox{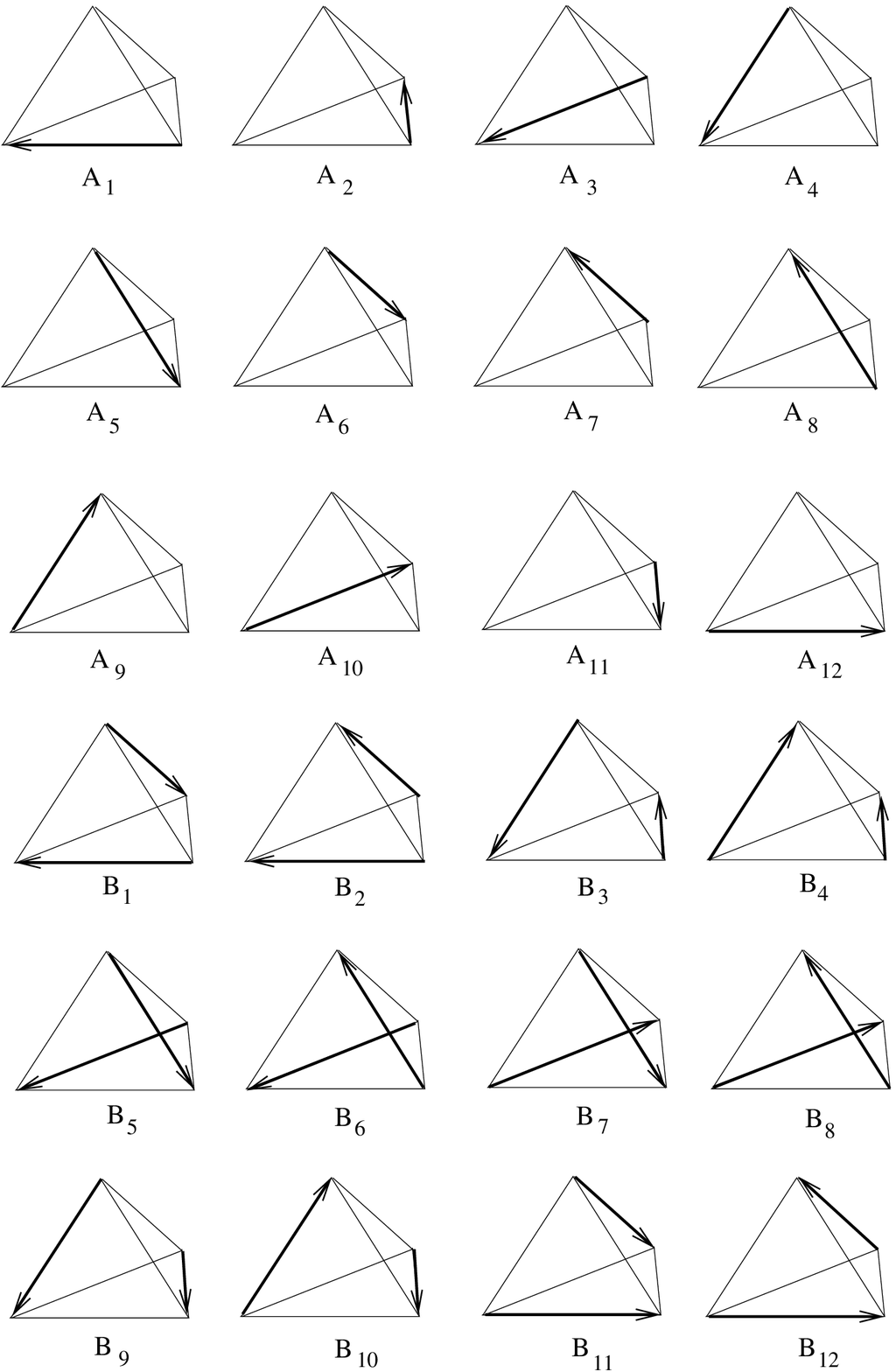}
\end{center}
\caption{In this figure we show the generating functions
at the first order together
with their names given in the text.}
\end{figure}

\begin{figure}
\begin{center}
\epsfxsize=5.in
\centerline{\epsfbox{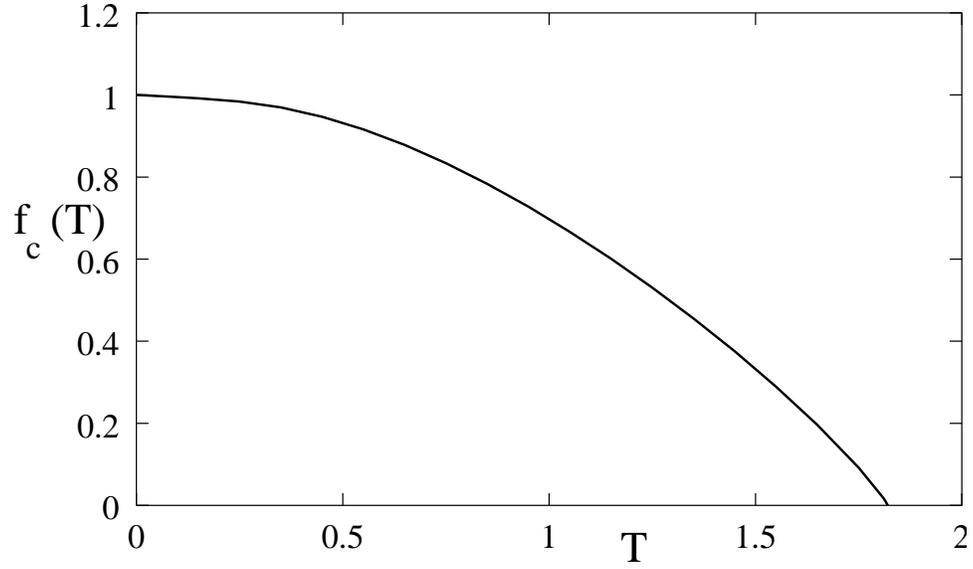}}
\end{center}
\caption{Plot of the phase diagram on the 3DSG found numerically. In this 
figure we have taken $\epsilon=1/2$ to make the calculations.}
\end{figure}

\begin{figure}
\begin{center}
\epsfxsize=5.in
\centerline{\epsfbox{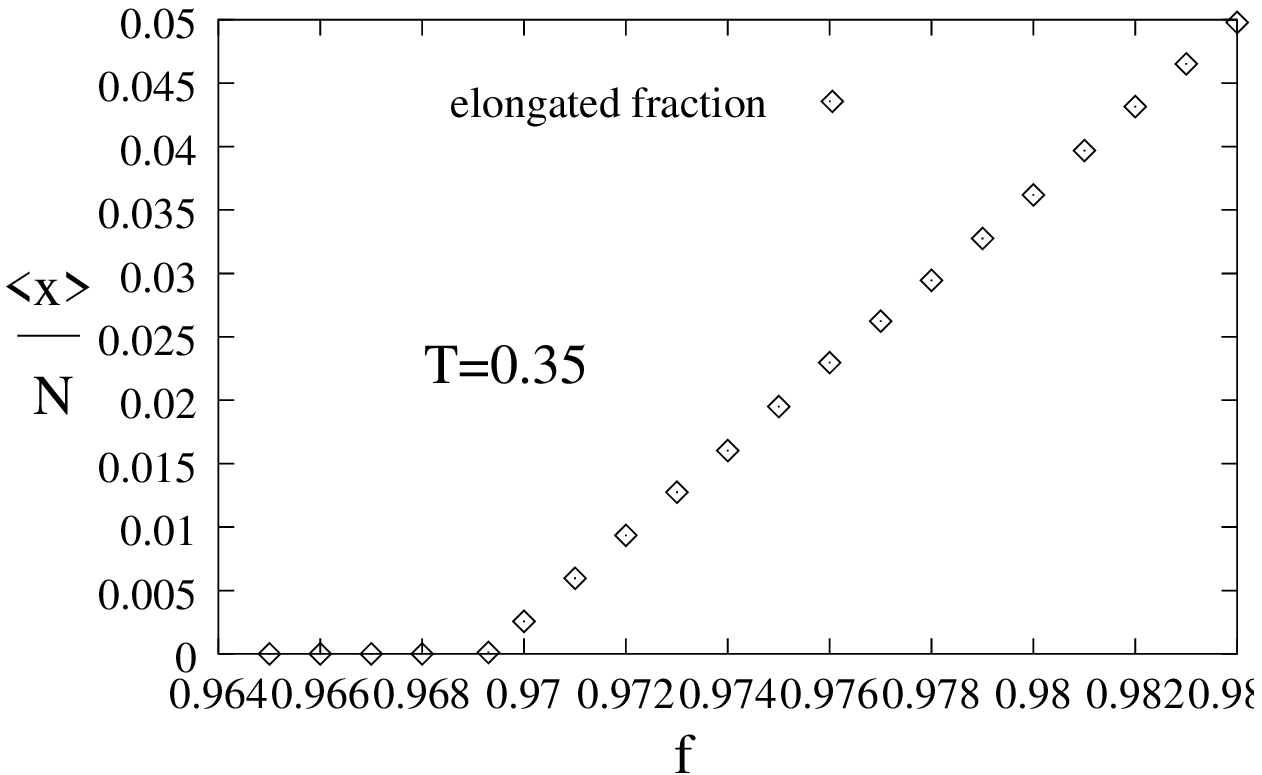}}
\end{center}
\caption{ Plot of the average elongation scaled by $N$
in the thermodynamic limit as a function of $f$ for $T=0.35$.
The critical force found numerically by imposing that the fixed point
reached after iteration of the recursion is that of Eq. \ref{critical_force}
is approximately $0.969\ldots$. We have taken $\epsilon=1/2$ in these
calculations.}
\end{figure}

\begin{figure}
\begin{center}
\epsfxsize=6.4in
\epsfbox{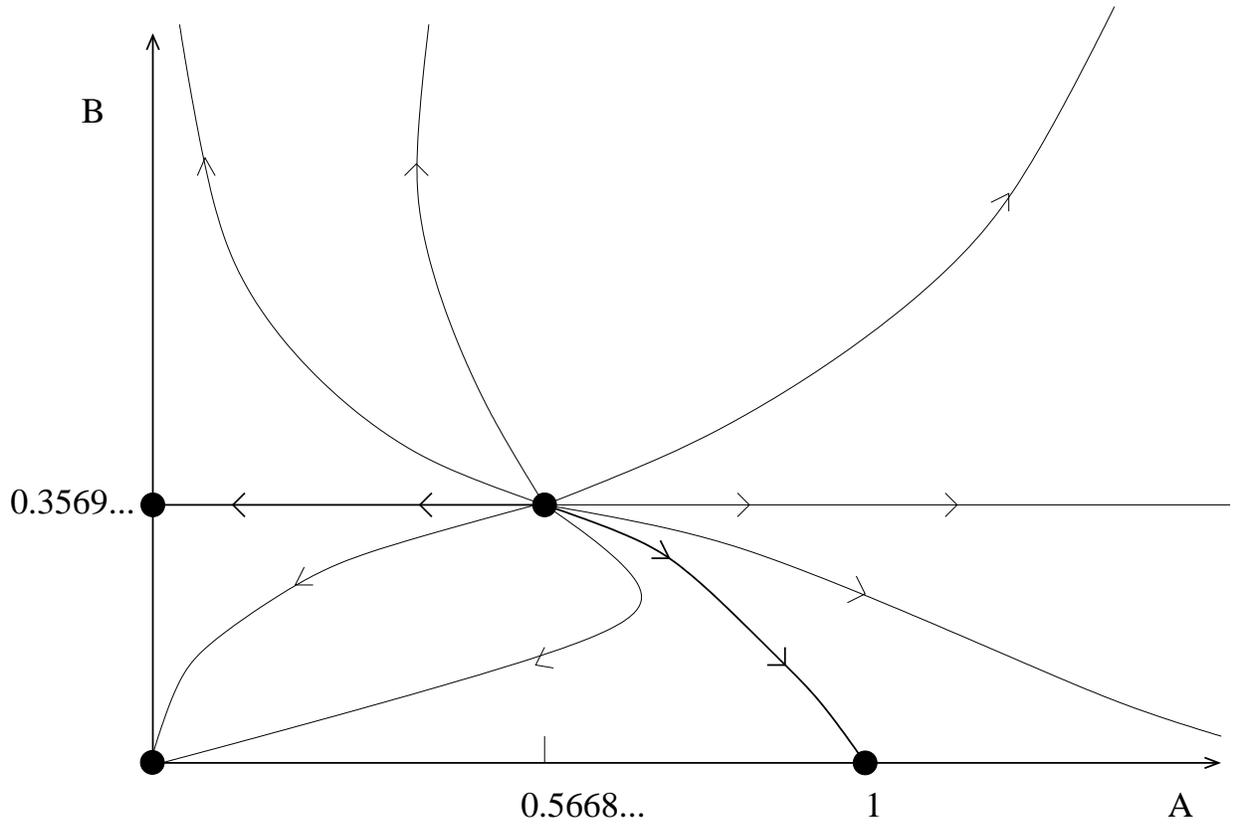}
\end{center}
\caption{Plot of the flux lines resulting from the simplified recursion 
relations corresponding to Eq. \ref{rg_approx}.}
\end{figure}

\end{document}